\documentclass[12pt]{article}
\usepackage{graphicx}
\usepackage{slashed}
\newcommand{\be}{\begin{eqnarray}}
\newcommand{\ee}{\end{eqnarray}}
\usepackage[colorlinks]{hyperref}
\hypersetup{
    colorlinks=true,
    citecolor=blue,
    filecolor=magenta,      
    urlcolor=cyan,
}

\usepackage{amsmath,amssymb}
\usepackage{mathtools}
\usepackage{cite}
\usepackage{float}
\usepackage{color}
\usepackage{booktabs}
\usepackage{subcaption}
\usepackage{amsfonts}
\usepackage[utf8]{inputenc}
\newcommand{\mathsym}[1]{{}}
\newcommand{\unicode}[1]{{}}

\begin{document}

\allowdisplaybreaks
\begin{center}

{\Large \boldmath \bf Probing anomalous gauge-Higgs couplings using $Z$ boson polarization at $e^+ e^-$ colliders} 

\vskip .4cm

{\large
Kumar Rao$^a$, Saurabh D. Rindani$^b$ and Priyanka Sarmah$^a$}\\
\vskip .2cm
{$^a$\it Physics Department, Indian Institute of Technology Bombay, \\Powai,
Mumbai 400076, India}\\
\vskip .2cm
{$^b$\it Theoretical Physics Division, Physical Research Laboratory,
\\Navrangpura, Ahmedabad 380009, India} \\

\vskip 1cm
{\bf Abstract}
\end{center}
We study possible new physics interactions in the $ZZH$ vertex contributing to the Higgsstrahlung process $e^+ e^- \to ZH$ at proposed future $e^+e^-$ colliders using the polarization of the $Z$ as a probe. We calculate the spin density matrix of the $Z$ for the process and determine the eight independent polarization parameters of the $Z$ boson which have the potential to constrain the anomalous couplings. We study angular asymmetries using the decay leptons from the $Z$ boson which are simply related to the polarization observables. We also estimate the limits that can be placed on the anomalous couplings using measurements of these angular asymmetries at centre of mass energies of 250 GeV and 500 GeV and various combinations of polarized $e^+$ and $e^-$ beams.

\section{Introduction}\label{sec1}
The discovery of the Higgs boson ($H$) with mass around 125 GeV at the Large Hadron Collider (LHC) completes the particle spectrum of the Standard Model (SM). However, to confirm that this is indeed the Higgs boson of the SM and uncover the exact mechanism of the breaking of the electroweak symmetry will require precise measurements of the couplings of the Higgs to electroweak gauge bosons ($V=W^\pm,Z,\gamma$), its Yukawa couplings to the fermions as well as its self couplings.  The $VVH$ coupling is of particular importance whose form is fixed by the $SU(2)_L \times U(1)_Y$ gauge structure of the SM.

Although the present accuracy of 
experiments at the LHC seems to indicate that the couplings of the Higgs boson are in broad agreement with the SM predictions, given their importance in confirming the symmetry breaking mechanism in the SM, one could be  optimistic that future experiments with higher luminosities will either be able to constrain such couplings even further or indicate small deviations from SM predictions. To do that one would need not just the cross section, 
but more observables. A fit to the differential cross section would be highly demanding requiring large statistics. However, use can be made of expectation values of  kinematic 
variables, or their asymmetries.

A number of scenarios going beyond the SM predict new particles and interactions at the TeV scale and require an enhanced Higgs sector resulting in modified couplings of the Higgs boson to SM particles. 
Various studies have been carried out on the $VVH$ coupling, at planned $e^+e^-$ colliders and the LHC, using a most general tensorial form of the
coupling and using a variety of observables involving kinematic distributions of the $Z$ and the charged leptons from $Z$ decay, see for example, \cite{Hagiwara:1993sw, Hagiwara:2000tk, Biswal:2005fh, Godbole:2007cn, Biswal:2008tg, Biswal:2009ar, Rindani:2009pb, Rindani:2010pi, Anderson:2013afp, Godbole:2014cfa, Godbole:2013lna, Craig:2015wwr, Beneke:2014sba, Khanpour:2017cfq, Zagoskin:2018wdo, Li:2019evl, He:2019kgh}. Anomalous interactions in the $HVV$ vertex have been searched for by the CMS collaboration \cite{Sirunyan:2019nbs, Sirunyan:2019twz} and although the current data are consistent with the SM predictions, the constraints are still weak (some details follow later) 
to allow for beyond the SM contributions to the vertex. 

An interesting variable which has received attention in recent
years is the polarization of the $Z$ boson produced in association with the Higgs. 
In this work, we study the $ZZH$ coupling using the associated production of the $Z$ with the Higgs at proposed future $e^+e^-$ colliders.
To measure polarization, one 
needs to look at kinematic distributions of some decay products.
In this work we adopt the formalism where we construct appropriate asymmetries from the angular 
distribution of the decay products of the $Z$, which in turn are related to 
its  polarization parameters  \cite{Aguilar-Saavedra:2017zkn, Boudjema:2009fz}. 
Measurement of these polarization parameters will give insight into the 
production mechanism and also provide  information about the  nature of
the tensorial structure as well as the strength of anomalous couplings in 
the $ZZH$ interaction. Such measurements can be used 
to place stringent limits on the anomalous couplings in the absence of any deviation 
from the SM prediction.

$Z$ polarization has been studied in the context of new physics at the
LHC \cite{Aguilar-Saavedra:2017zkn, Nakamura:2017ihk}. 
Refs. \cite{Rahaman:2017qql, Rahaman:2016pqj} studied anomalous coupling in 
$ZZ$/$Z\gamma$ production at an $e^{-}e^{+}$ collider. Ref. \cite{Rindani:2010pi} studied 
various general
$ZZH$ and $Z\gamma H$ couplings using a different set of 
observables. Analogously, polarization of the $W$ boson produced in
association with the Higgs at the LHC has been
studied in \cite{Rao:2018abz, Nakamura:2017ihk}.

In this paper, we are addressing the question how well the form 
factors for the $ZZH$ interaction can be determined from the polarization 
observables of the $Z$ boson produced in association with the Higgs
boson at an $e^+e^-$ collider. There have been several proposals for
$e^+e^-$ colliders, especially to be employed as ``Higgs factories",
operating around 240-250 GeV centre-of-mass (c.m.) energy. These are the
International Linear Collider (ILC) in Japan \cite{Asner}, the Circular Electron
Positron Collider (CEPC) in China \cite{CEPC}, 
the Future Circular Collider with
$e^+e^-$ (FCC-ee) at CERN (previously known as TLEP \cite{TLEP}). 
The Compact Linear
Collider (CLIC) at CERN would possibly run at higher c.m. energies
\cite{CLIC:2016zwp}.
These colliders are planned to collect different integrated luminosities
over time, which could be as high as 10 ab$^{-1}$ as for example in the
case of the FCC-ee. We restrict ourselves to the illustrative case of
the ILC operating at 250 GeV collecting 2 ab$^{-1}$ and at 500 GeV at a
later stage, with 500 fb$^{-1}$ of data. 
It may be borne in mind that the sensitivities that we estimate may be
conservative, and may be better at another collider with a larger
luminosity.

We include in our discussion  unpolarized
beams as well as longitudinally polarized beams which are likely to be available at the
ILC. Beam polarization has been known to
have the advantage of suppressing certain backgrounds and enhancing the
sensitivity by appropriate choice of signs of the polarization (see, for
example, \cite{MoortgatPick:2005cw}).

We therefore make predictions for
the $Z$ polarization variables, as also certain kinematic asymmetries in
leptonic variables, since charged leptonic decay modes of the $Z$ would be ideal
for the study of the polarization. It is expected that the ILC would
initially run at a c.m. energy of 250 GeV, and that there would be a
later run with c.m. energy 500 GeV. We consider both these
possibilities. We also estimate the sensitivity of the leptonic
asymmetries to possible $ZZH$ anomalous couplings, under ideal
experimental conditions.

We consider the process $e^{-}e^{+}\rightarrow Z H$,
where  the vertex $Z_{\mu}(k_{1})\rightarrow Z_{\nu}(k_{2}) H$  has the 
Lorentz structure
\begin{equation}\label{vertex}
 \Gamma^{V}_{\mu \nu} =\frac{g}{\cos\theta_{W}}m_{Z} \left[ a_{Z}g_{\mu 
\nu}+
\frac{ b_{Z}}{m_{Z}^{2}}\left( k_{1 \nu}k_{2 \mu}-g_{\mu \nu}  k_{1}. 
k_{2}\right) +\frac{\tilde b_{Z}}{m_{Z}^{2}}\epsilon_{\mu \nu \alpha \beta} 
k_{1}^{\alpha} k_{2}^{\beta}\right]  
 \end{equation}
where $g$ is the $SU(2)_L$ coupling and $\theta_{W}$ is the weak mixing angle. The form factors $ a_{Z}$, $b_{Z}$ and $\tilde b_{Z}$ are in general complex.
 The first two couplings would correspond to CP-even terms in the interaction, 
while the third term is odd under CP. In the SM, the coupling $ a_{Z}$
is unity at tree level, whereas the other two couplings $b_{Z}$ , $\tilde b_{Z}$ 
vanish at tree level, denoting the deviation from the tree-level SM value. 
Such anomalous couplings  could arise from loop corrections in the SM or in any 
extension of SM with some new particles. However, we are not concerned with the 
latter and derive the helicity amplitudes for the process of our interest in a 
model-independent way, which follows in the next section, where we also
evaluate the spin density matrix, which is crucial for a complete
description of polarization.

At the LHC, both ATLAS and CMS collaborations have attempted to constrain 
the $ZZH$ anomalous couplings, though the limits are not very stringent. 
For example, CMS has put bounds on ratios of the cross section contributions
arising from the different $ZZH$  couplings 
\cite{Sirunyan:2019nbs}
which in our notation translate 
to $\vert {\rm Re }b_Z/a_Z \vert < 0.058$ and $\vert {\rm Re }\tilde b_Z/
a_Z \vert < 0.078 $ .
Similarly, ATLAS \cite{ATLAS} has put bounds on coefficients of an effective 
Lagrangian, assumed 
nonzero one at a time, which are related to the anomalous couplings we discuss.
The bounds are again of about a few per cent. The possibility of
a future Large Hadron electron Collider (LHeC) to probe anomalous $ZZH$ 
couplings has been studied in \cite{Cakir:2013bxa}, where weak limits are
found, viz.,   $-0.21 < b_Z <  0.43$ and  $-0.32 < \tilde{b}_Z < 0.32$ 
for an electron beam energy of 60 GeV and mild improvement for a beam energy of 140 GeV, with proton beam energy of 7 TeV in either case.
At $e^+e^-$ colliders, 
projections have been made for the measurement of these anomalous couplings 
in the context of various planned machines. Apart from measurement of 
the cross section of the dominant process $e^+e^- \to HZ$, 
suggestions have been made to include decay distributions as well as results of 
HL-LHC (see
\cite{futureee} for some relevant references)
in a global analysis of various effective Lagrangian parameters.
The estimates of sensitivity indicate (see for example 
\cite{Durieux:2017rsg}) that while
HL-LHC would provide limits of the order of order $10^{-1}$ or $10^{-2}$, these would be improved to a per mille level on using angular distributions at electron-positron colliders. Our projections for $HZZ$ couplings are
consistent with this general expectation, at the same time, providing a physical understanding through $Z$ polarization parameters.
 
\section{Helicity amplitudes and density matrix}\label{sec2}
We compute the helicity amplitudes for the process 
\begin{equation}
 e^{-}(p_{1}) +e^{+}(p_{2}) \rightarrow Z^{\alpha}(p) + H(k)
\end{equation}
in the massless limit of the initial particles, with the $ZZH$ vertex given in Eqn.(\ref{vertex}). We will later construct angular asymmetries using the charged muon from $Z$ decay. This process receives a contribution from the ``Higgsstrahlung'' diagram, mediated by a $s$-channel $Z$. Our results also hold for the $Z$ decaying to taus, to the extent that the mass of the taus can be neglected. We do not consider $Z$ decay to $e^+e^-$ to avoid interference from the SM vector boson fusion diagram,  though these effects are numerically small.

To calculate these amplitudes we adopt the following representations for the 
transverse and longitudinal polarization vectors of the $Z$:
 \begin{equation}
 \epsilon^{\mu}(p,\pm)=\frac{1}{\sqrt{2}}(0,\mp \cos\theta, - i ,\pm \sin\theta),
  \end{equation}
   \begin{equation}
   \epsilon^{\mu}(p,0)=\frac{1}{m_{Z}}(\vert \vec{p}_{Z}\vert, E_{Z}\sin\theta,0,E_{Z}\cos\theta),
 \end{equation}
 where $E_{Z}$ and $\vec{p}_{Z}$ are the energy and momentum of the $Z$ 
respectively, with $\theta$ being the polar angle made by the $Z$ with respect to the $e^{-}$ momentum taken to be  along the positive $z$ axis.

The helicity amplitudes which are non-zero in the limit of massless initial 
states are
\begin{eqnarray}
M(-,+,+)&=&\frac{g^{2}m_{Z}\sqrt{s}(c_{V}+c_{A})}
{2\sqrt{2}\cos^{2}\theta_{W}(s-m_{Z}^{2})}\left[  1- \frac{\sqrt{s}}{ m_{Z}^{2}}(E_{Z}b_{z}+ i  \tilde b_{z}\vert \vec{p}_{Z}\vert)
\right]\\ \nonumber
&& \times  (1-\cos\theta)\\
M(-,+,-)&=&\frac{g^{2}m_{Z}\sqrt{s}(c_{V}+c_{A})}
{2\sqrt{2}\cos^{2}\theta_{W}(s-m_{Z}^{2})}\left[  1- \frac{\sqrt{s}}{ m_{Z}^{2}}(E_{Z}b_{Z}- i  \tilde b_{Z}\vert \vec{p}_{Z}\vert)
\right] \\ \nonumber
&& \times (1+\cos\theta)\\
M(-,+,0)&=&\frac{g^{2}\sqrt{s}(c_{V}+c_{A})}
{2\cos^{2}\theta_{W}(s-m_{Z}^{2})}\left[E_{Z}-\sqrt{s}b_{Z}\right]  \sin\theta \\
M(+,-,+)&=&\frac{-g^{2}m_{Z}\sqrt{s}(c_{V}-c_{A})}
{2\sqrt{2}\cos^{2}\theta_{W}(s-m_{Z}^{2})}\left[  1-
  \frac{\sqrt{s}}{ m_{Z}^{2}}(E_{Z}b_{Z}+ i  \tilde b_{Z} \vert \vec{p}_{Z}\vert)\right] \\ \nonumber 
  && \times (1+\cos\theta)\\
M(+,-,-)&=&\frac{-g^{2}m_{Z}\sqrt{s}(c_{V}-c_{A})}
{2\sqrt{2}\cos^{2}\theta_{W}(s-m_{Z}^{2})}\left[  1-
  \frac{\sqrt{s}}{ m_{Z}^{2}}(E_{Z}b_{Z}- i  \tilde b_{Z}\vert \vec{p}_{Z}\vert)\right] \\ \nonumber
  && \times (1-\cos\theta)\\
M(+,-,0)&=&\frac{g^{2}\sqrt{s}(c_{V}-c_{A})}{
2\cos^{2}\theta_{W}(s-m_{Z}^{2})}\left[E_{Z}-\sqrt{s}b_{Z}\right]  \sin\theta
\end{eqnarray}
Here the first two entries in $M$ denote the signs of the helicities of 
the electron and positron respectively and the third entry is the $Z$ helicity. $\sqrt{s}$ is
the total c.m. energy, and  $c_{V}=-\frac{1}{2}+2\sin^{2}\theta_{W}$, 
$c_{A}=-\frac{1}{2}$ are the vector and axial vector couplings of the $Z$ 
to charged leptons.
 
In deriving these helicity amplitudes, we have assummed the SM value 
$a_Z=1$ for $a_Z$. 
However, the $a_Z$ dependence can be easily recovered
by multiplying the helicity amplitude expressions by $a_Z$,
and then replacing $b_Z$ and $\tilde b_Z$ by $b_Z/a_Z$ and $\tilde b_Z/a_Z$, 
respectively. 

The spin-density matrix for $Z$ production expressed in terms of the helicity amplitudes is given by
 \begin{equation}\label{rhodef}
 \rho(i,j)=\overline\sum_{\lambda,\lambda^{'}}M(\lambda,\lambda^{'},i)M^{\ast}(\lambda,\lambda^{'},j)
 \end{equation}
 the average being over the initial helicities $\lambda$, $\lambda^{'}$ 
of the electron and positron respectively and the indices $i,j$ can take values $\pm,0$.
The diagonal elements of  Eqn.(\ref{rhodef}) with $i=j$ would give the production probabilities of 
$Z$ with definite polarization, whose ratios to the total cross section are 
known as helicity fractions for the corresponding polarizations. 
Apart from  these  diagonal elements,
it is also necessary to know the off-diagonal elements 
to include the spin information in a coherent way in the combination of the 
$Z$ production and decay processes.  
Then, on integrating over the  phase 
space, the full density matrix Eqn.(\ref{rhodef}) would lead to the eight independent 
vector and tensor polarization components, known as the \textit{polarization 
parameters} of the $Z$. 

The density matrix elements, for unpolarized $e^+$ and $e^-$ beams, derived from the helicity amplitudes, to linear 
order in the anomalous couplings $b_{Z}$ and $\tilde b_{Z}$ and setting $a_{Z}=1$ are given by
\begin{eqnarray}
 \rho(\pm,\pm
)&=&\frac{g^{4}m^{2}_{Z}s}{8\cos^{4}\theta_{W}(s-m_{Z}^{2})^2}
\left[
(c_{V}+c_{A})^{2}(1\mp\cos\theta)^{2}\right. \nonumber\\
&& \hskip -0.8cm \left. +(c_{V}-c_{A})^{2}(1\pm\cos\theta)^{2}\right]
\left[1-2(\text{Re}~b_{Z}\mp\beta_{Z}\text{Im}~\tilde b_{Z})\frac{
E_{Z}\sqrt{s}}{m^{2}_{Z}} \right] \\
\rho(0,0
)&=&\frac{g^{4}E^{2}_{Z}s}{2\cos^{4}\theta_{W}(s-m_{Z}^{2})^2}\sin^{2}\theta
\, (c_V^2 + c_A^2) \left[1-2{\rm Re}~b_{Z}\frac{\sqrt{s}}{E_{Z}}\right] \\
\rho(\pm,\mp
)&=&\frac{g^{4}m^{2}_{Z}s}{4\cos^{4}\theta_{W}(s-m_{Z}^{2})^2}\sin^{2}\theta
 \,(c_V^2 + c_A^2)\nonumber \\ &&\times 
\left[1-2({\rm Re}~b_{Z}\pm  i \beta_{Z} {\rm Re}~\tilde b_{Z})\frac{
E_{Z}\sqrt{s}}{m^{2}_{Z}} \right] \\ \nonumber
\rho(\pm,0 )&=&\frac{g^{4}m_{Z}E_{Z}s}
{4\sqrt{2}\cos^{4}\theta_{W}(s-m_{Z}^{2})^2}\sin\theta
\nonumber \\
&& \hskip -.8cm \times \left[
(c_{V}+c_{A})^{2}(1\mp\cos\theta) 
 -(c_{V}-c_{A})^{2}(1\pm\cos\theta)\right] \nonumber\\
&& \hskip -.8cm \times 
\left[1-{\rm Re}~b_{Z}\sqrt{s}\frac{
(E^{2}_{Z}+m^{2}_{Z})}{E_{Z}m^{2}_{Z}}
 - i \sqrt{s} \frac{E_{Z}}{m^{2}_{Z}}\left({\rm
Im}~b_{Z}\beta^{2}_{Z}\pm \tilde
b_{Z} \beta_{Z})\right)\right] 
\end{eqnarray}
where $\beta_{Z}=\vert\vec{p}_{Z}\vert /E_{Z}$ is the velocity of the $Z$ in 
the c.m frame. The analytical manipulation software FORM 
\cite{Vermaseren:2000nd} has been used to verify these expressions.
We have kept the finite $Z$ width 
in our numerical calculations later.

The density matrix elements in Eqns. (12)-(15) are
computed to leading order in the anomalous couplings, by taking the
overlap of the BSM amplitudes with the SM amplitude for which $a_Z = 1$. Any
value of $a_Z$ not equal to one would represent BSM physics and thus its
overlap with the $b_Z$ and $\tilde{b}_Z$ pieces would be quadratic in the
anomalous couplings, which we neglect, assuming them to be small.

We do not display the somewhat longer expressions for the density matrix 
elements taking into account the polarizations $P_L$ and $\bar{P}_L$ of the 
electron and positron beams, respectively. However, the expressions are more 
compact on integration over $\cos \theta$, and these are displayed here:
\begin{eqnarray}
 \sigma(\pm,\pm
)&=&\frac{2(1-P_L\bar P_L)g^{4}m^{2}_{Z}s}{3\cos^{4}\theta_{W}(s-m_{Z}^{2})^2}
(c_{V}^2+c_{A}^{2}-2P_L^{\rm eff}c_Vc_A) \nonumber\\
&& \hskip -0.8cm 
\times\left[1-2(\text{Re}~b_{Z}\mp\beta_{Z}\text{Im}~\tilde b_{Z})\frac{
E_{Z}\sqrt{s}}{m^{2}_{Z}} \right] \\
\sigma(0,0
)&=&\frac{2(1-P_L\bar P_L)g^{4}E^{2}_{Z}s}{3\cos^{4}\theta_{W}(s-m_{Z}^{2})^2}
(c_V^2 + c_A^2 -2P_L^{\rm eff}c_Vc_A)\nonumber \\ 
&& \hskip -0.8cm
\times \left[1-2{\rm Re}~b_{Z}\frac{\sqrt{s}}{E_{Z}}\right] \\
\sigma(\pm,\mp
)&=&\frac{(1-P_L\bar P_L^{\rm eff})g^{4}m^{2}_{Z}s}{3\cos^{4}\theta_{W}(s-m_{Z}^{2})^2}
 \,(c_V^2 + c_A^2 - 2P_L^{\rm eff}c_Vc_A)\nonumber \\ 
&& \hskip -0.8cm
\times \left[1-2({\rm Re}~b_{Z}\pm  i \beta_{Z} {\rm Re}~\tilde b_{Z})\frac{
E_{Z}\sqrt{s}}{m^{2}_{Z}} \right] \\ \nonumber
\sigma(\pm,0 )&=&\frac{(1-P_L\bar P_L^{\rm eff})\pi g^{4}m_{Z}E_{Z}s}
{4\sqrt{2}\cos^{4}\theta_{W}(s-m_{Z}^{2})^2}
\left[ (2c_{V}c_{A} - P_L^{\rm eff}(c_V^2+c_A^2))
 \right] \nonumber\\
&& \hskip -.8cm \times 
\left[1-{\rm Re}~b_{Z}\sqrt{s}\frac{
(E^{2}_{Z}+m^{2}_{Z})}{E_{Z}m^{2}_{Z}}
 - i \sqrt{s} \frac{E_{Z}}{m^{2}_{Z}}\left({\rm
Im}~b_{Z}\beta^{2}_{Z}\pm \tilde
b_{Z} \beta_{Z})\right)\right] 
\end{eqnarray}
In the above equations, $P_L^{\rm eff} = (P_L - \bar{P}_L)/ 
( 1 - P_L \bar{P}_L)$.

It is interesting to note that in the density matrix elements, while the SM
contributions show the typical dominance of longitudinal polarization as 
as compared to the transverse polarizations, the contributions linear in $b_Z$ 
do not show this behaviour. The reason for this is the form of the tensor 
in the $b_Z$ vertex in (1), for which the leading contribution from the 
longitudinal polarization vector of the $Z$ vanishes. In fact, as will be seen 
later, the longitudinal and transverse contributions to the $b_Z$ dependence
of the cross section turn out to be numerically equal.  

Defining an integral of this density matrix over an appropriate kinematic
range as $\sigma(i,j)$, the latter can be parametrized in terms of the linear 
polarization $\vec P$ and the tensor polarization $T$ as follows \cite{Leader:2001gr}.
\begin{equation}\label{vectensorpol}
\sigma(i,j) \equiv \sigma \;\left( 
\begin{array}{ccc}
\frac{1}{3} + \frac{P_z}{2} + \frac{T_{zz}}{\sqrt{6}} &
\frac{P_x - i P_y}{2\sqrt{2}} + \frac{T_{xz}-i T_{yz}}{\sqrt{3}} &
\frac{T_{xx}-T_{yy}-2iT_{xy}}{\sqrt{6}} \\
\frac{P_x + i P_y}{2\sqrt{2}} + \frac{T_{xz}+i T_{yz}}{\sqrt{3}} &
\frac{1}{3} - \frac{2 T_{zz}}{\sqrt{6}} &
\frac{P_x - i P_y}{2\sqrt{2}} - \frac{T_{xz}-i T_{yz}}{\sqrt{3}} \\
\frac{T_{xx}-T_{yy}+2iT_{xy}}{\sqrt{6}} &
\frac{P_x + i P_y}{2\sqrt{2}} - \frac{T_{xz}+i T_{yz}}{\sqrt{3}} &
\frac{1}{3} - \frac{P_z}{2} + \frac{T_{zz}}{\sqrt{6}} 
\end{array}
\right)
\end{equation}
where $\sigma(i,j)$ is the integral of $\rho(i,j)$, and $\sigma$ is the 
production cross section, 
\begin{equation}
\sigma = \sigma(+,+) + \sigma(-,-) + \sigma(0,0).
\end{equation}
In the following section, we construct the polarization parameters from
the integrated density matrix elements and give relations of these to 
angular asymmetries of the decay leptons which would serve as measures
of the polarization parameters. 

\section{$Z$ polarization and lepton asymmetries}\label{sec3}
The eight independent vector and tensor polarization observables can be
extracted using appropriate linear combinations of the integrated
density matrix elements of Eqn.(\ref{vectensorpol}):
 \begin{eqnarray}\label{pol1}
P_{x}&=&  \frac{\lbrace \sigma(+,0)+\sigma(0,+)\rbrace
 +\lbrace \sigma(0,-)+\sigma(-,0)\rbrace}{\sqrt{2}\sigma}\\\label{pol2}
 P_{y}&=&\frac{- i  \lbrace[\sigma(0,+)-\sigma(+,0)]+[\sigma(-,0)-\sigma(0,-)]\rbrace}{\sqrt{2}\sigma}\\\label{pol3}
 P_{z}&=&\frac{[\sigma(+,+)]-[\sigma(-,-)]}{\sigma}\\\label{pol4}
 T_{xy}&=&\frac{- i  \sqrt{6}[\sigma(-,+)-\sigma(+,-)]}{4\sigma}\\\label{pol5}
  T_{xz}&=&\frac{\sqrt{3}\lbrace[\sigma(+,0)+\sigma(0,+)]-[\sigma(0,-)+\sigma(-,0)]\rbrace}{4\sigma}\\\label{pol6}
  T_{yz}&=&\frac{- i  \sqrt{3}\lbrace[\sigma(0,+)-\sigma(+,0)]-[\sigma(-,0)-\sigma(0,-)]\rbrace}{4\sigma}\\\label{pol7}
  T_{xx}-T_{yy}&=&\frac{\sqrt{6}[\sigma(-,+)+\sigma(+,-)]}{2\sigma}\\\label{pol8}
  T_{zz}&=&\frac{\sqrt{6}}{2}\left\lbrace\frac{[\sigma(+,+)]+[\sigma(-,-)]}{\sigma}-\frac{2}{3}\right\rbrace
\nonumber \\
 &=&\frac{\sqrt{6}}{2}\left[\frac{1}{3}-\frac{\sigma(0,0)}{\sigma}\right]\label{pol9}
 \end{eqnarray}
Of these $P_x$, $P_y$ and $P_z$ are the vector polarizations, whereas
the $T$'s are the tensor polarizations, with the constraint that the
tensor is traceless.

Experimentally, the spin information of the $Z$ is obtained from kinematic
distributions of its decay products. 
Ref. \cite{Rahaman:2017qql} describes
  the formalism that connects 
all the spin observables of $Z$  to the angular distribution of the leptons 
arising from its decay. Ref. \cite{Rahaman:2017qql} also defines
 various asymmetries 
in the rest frame of the $Z$ which are simply related to
the polarization observables given in Eqs.(\ref{pol1})-(\ref{pol9}). These 
are given by
\begin{equation}\label{asyx}
A_{x}=\frac{3\alpha P_{x}}{4}\equiv\frac{\sigma(\cos\phi^{\ast}>0)-\sigma(\cos\phi^{\ast}<0)}{\sigma(\cos\phi^{\ast}>0)+\sigma(\cos\phi^{\ast}<0)}
\end{equation}
\begin{equation}\label{asyy}
A_{y}=\frac{3\alpha P_{y}}{4}\equiv\frac{\sigma(\sin\phi^{\ast}>0)-\sigma(\sin\phi^{\ast}<0)}{\sigma(\sin\phi^{\ast}>0)+\sigma(\sin\phi^{\ast}<0)}
\end{equation}
\begin{equation}\label{asyz}
A_{z}=\frac{3\alpha P_{z}}{4}\equiv\frac{\sigma(\cos\theta^{\ast}>0)-\sigma(\cos\theta^{\ast}<0)}{\sigma(\cos\theta^{\ast}>0)+\sigma(\cos\theta^{\ast}<0)}
\end{equation}
\begin{equation}\label{asyxy}
A_{xy}=\frac{2}{\pi}\sqrt{\frac{2}{3}}T_{xy}\equiv\frac{\sigma(\sin2\phi^{\ast}>0)-\sigma(\sin2\phi^{\ast}<0)}{\sigma(\sin2\phi^{\ast}>0)+\sigma(\sin2\phi^{\ast}<0)}
\end{equation}
\begin{equation}\label{asyxz}
A_{xz}=\frac{-2}{\pi}\sqrt{\frac{2}{3}}T_{xz}\equiv\frac{\sigma(\cos\theta^{\ast}\cos\phi^{\ast}<0)-\sigma(\cos\theta^{\ast}\cos\phi^{\ast}>0)}{\sigma(\cos\theta^{\ast}\cos\phi^{\ast}>0)+\sigma(\cos\theta^{\ast}\cos\phi^{\ast}<0)}
\end{equation}
\begin{equation}\label{asyyz}
A_{yz}=\frac{2}{\pi}\sqrt{\frac{2}{3}}T_{yz}\equiv\frac{\sigma(\cos\theta^{\ast}\sin\phi^{\ast}>0)-\sigma(\cos\theta^{\ast}\sin\phi^{\ast}<0)}{\sigma(\cos\theta^{\ast}\sin\phi^{\ast}>0)+\sigma(\cos\theta^{\ast}\sin\phi^{\ast}<0)}
\end{equation}
\begin{equation}\label{asyx2y2}
A_{x^{2}-y^{2}}=\frac{1}{\pi}\sqrt{\frac{2}{3}}(T_{xx}-T_{yy})\equiv\frac{\sigma(\cos2\phi^{\ast}>0)-\sigma(\cos2\phi^{\ast}<0)}{\sigma(\cos2\phi^{\ast}>0)+\sigma(\cos2\phi^{\ast}<0)}
\end{equation}
\begin{equation}\label{asyzz}
A_{zz}=\frac{3}{8}\sqrt{\frac{3}{2}}T_{zz}\equiv\frac{\sigma(\sin3\theta^{\ast}>0)-\sigma(\sin3\theta^{\ast}<0)}{\sigma((\sin3\theta^{\ast}>0)+\sigma((\sin3\theta^{\ast}<0)}
\end{equation}
Here, $\alpha$ is the $Z$ boson polarization analyser, given in terms of its left and right handed couplings to charged leptons, $L_\ell$ and $R_\ell$ respectively, as
\begin{equation}\label{polanalyser}
\alpha =\frac{R_\ell^2 -L_\ell ^2}{R_\ell^2 +L_\ell ^2}=-\frac{2 c_V c_A}{c_V^2 +c_A^2}
\end{equation}
The angles $\theta^\ast$ and $\phi^\ast$ are polar and azimuthal angles of
the lepton in the rest frame of the $Z$. This frame is reached by a
boost from the laboratory frame. In the laboratory frame, the initial
$e^-$ beam defines the $z$ axis, and the production plane of $Z$ is
defined as the $xz$. While boosting to the $Z$ rest frame, the $xz$
plane is kept unchanged. Then, the angles $\theta^\ast$ and $\phi^\ast$
are  measured with respect to the
would-be momentum of the $Z$. 

We have evaluated these asymmetries in the case when initial beams are 
unpolarized or longitudinally polarized.
It is observed that out of eight polarizations asymmetries only 
three, {\it viz.}, $A_{x}$, $A_{x^{2}-y^{2}}$ and $A_{zz}$ are non-zero in the 
SM, which,  along 
with the total cross section, would be proportional to the real part of the 
anomalous couplings to satisfy CPT theorem. This will be seen in the following 
section.
Also, it can be seen from Eqs. (16)-(19) that only the asymmetries $A_x$, $A_y$,
$A_{xz}$ and $A_{yz}$, which involve $\sigma(\pm,0)$ or $\sigma(0,\pm)$ in their
definition through the corresponding polarization parameters have beam 
polarization dependence. The polarization dependence cancels between numerator
and denominator in the remaining asymmetries.

\section{Numerical results}
\label{sec4}

Here we present numerical values for the integrated density matrix
elements, the corresponding asymmetries and the sensitivities of the
asymmetries to the various anomalous 
couplings. We consider two possibilities for the collider parameters,
{\it viz.},
c.m. energy $\sqrt{s}=250\text{~GeV}$, with integrated luminosity
$\int \mathcal{L}dt=2$ $\text{ab}^{-1}$ 
and c.m. energy
$500\text{~GeV}$ with integrated 
luminosity 
$\int \mathcal{L}dt=500$ 
fb$^{-1}$. We consider longitudinal polarizations of $P_{L}=\pm 
0.8$ for the electron beam and  $\bar{P_{L}}=\pm 0.3$  for the
positron beam, which are expected to be available at the collider.
We assume, for simplicity, that the run with polarized beams is for
the same luminosity as the run with unpolarized beams.
We found that from among combinations with different signs of
polarizations, the polarization combination $(P_L,\bar P_L)=(-0.8,+0.3)$
corresponds to the best sensitivity for
given magnitudes of polarizations, and therefore we present results
only for this combination, in addition to those for unpolarized beams.

In Tables \ref{table:1} and \ref{table:2} we present for
c.m. energy 250 GeV the
numbers for the integrated production density matrix elements
$\sigma(i,j)$, $i,j$ taking values $\pm 1$ and $0$, respectively for
unpolarized beams and $(P_L,\bar P_L) = (-0.8,+0.3)$.
In each table, the column labelled SM lists the values for
the SM, whereas the other columns list the coefficients of the
respective couplings in the density matrix element.
The corresponding numbers for c.m. energy 500 GeV are given in Tables
\ref{table:3} and \ref{table:4}.
{
\renewcommand{\arraystretch}{1.3}
\begin{table}[H]
\centering
\begin{tabular}{|c|c|c|c|c|c|}
\multicolumn{5}{c}{} \\
 \hline
  & SM & $\text{Re}~b_{z}$  & $\text{Im}~b_{z}$ &
     $\text{Re}~\tilde{b}_{z}$&$ \text{Im}~\tilde{b}_{z}$   \\
\hline
$\sigma(\pm,\pm)$   & $ 70.32 $ &$-466.72 $ &$0$
& $0$ &  $\pm 262.99 $ \\
$\sigma(0,0)$       & $103.03$ &$-466.72 $&$0$ &  $0$ &
 $0$\\
$\sigma(\pm,\mp)$  & $35.16$   &$-233.36$ &$0$ &$\mp i131.49 $  &  $0$\\
$\sigma(\pm,0)$    & $ 10.60$ & $ -59.21$ & $-i11.17 $ &
$\mp i19.83  $ & $\pm19.83 $\\
$\sigma(0,\pm)$    & $ 10.60$ & $ -59.21$ & $ +i11.17 $ &
$\pm i19.83 $ & $\pm19.83 $\\
\hline
\end{tabular}
\caption{Production spin density matrix elements (in fb) of $Z$ for the
SM and the coefficients of various couplings in each matrix element for
unpolarized beams at $\sqrt{s}=250\text{~GeV}$.}
\label{table:1}
\end{table}}

{
\renewcommand{\arraystretch}{1.3}
\begin{table}[H]
\centering
\begin{tabular}{|c|c|c|c|c|c|}
 \multicolumn{5}{c}{} \\
 \hline
  & SM & $\text{Re}~b_{z}$  & $\text{Im}~b_{z}$ &
     $\text{Re}~\tilde{b}_{z}$&$ \text{Im}~\tilde{b}_{z}$   \\
\hline
$\sigma(\pm,\pm)$   & $ 98.76 $ &$-655.52 $ &$0$
& $0$ &  $\pm 369.38$ \\
$\sigma(0,0)$       & $144.71$ &$-655.52 $&$0$ &  $0$ &
 $0$\\
$\sigma(\pm,\mp)$  & $49.38$   &$-327.76$ &$0$ &$\mp i184.69 $  &  $0$\\
$\sigma(\pm,0)$    & $ 91.15$ & $ -508.93$ & $-i96.05 $ &
$\mp i170.45  $ & $\pm170.45 $\\
$\sigma(0,\pm)$    & $ 91.15$ & $ -508.93$ & $+ i96.05 $ &
$\pm i170.45 $ & $\pm170.45 $\\
\hline
\end{tabular}
\caption{Production spin density matrix elements (in fb) of $Z$ for the
SM and the coefficients of various couplings in each matrix element for
($P_{L}=-0.8,\bar{P}_{L}=0.3$)   at $\sqrt{s}=250\text{~GeV}$.}
\label{table:2}
\end{table}}

{
\renewcommand{\arraystretch}{1.3}
\begin{table}[H]
\centering
\begin{tabular}{|c|c|c|c|c|c|}
 \multicolumn{5}{c}{} \\
 \hline
  & SM & $\text{Re}~b_{z}$  & $\text{Im}~b_{z}$ &
     $\text{Re}~\tilde{b}_{z}$&$ \text{Im}~\tilde{b}_{z}$   \\
\hline
$\sigma(\pm,\pm)$   & $ 6.39 $ &$-186.51$ &$0$
& $0$ &  $\pm 172.85$ \\
$\sigma(0,0)$       & $45.26$ &$-186.51$&$0$ &  $0$ &
 $0$\\
$\sigma(\pm,\mp)$  & $3.19$   &$-93.26$ &$0$ &$\mp i86.42$  &  $0$\\
$\sigma(\pm,0)$    & $ 2.12$ & $ -35.29$ & $-i26.56 $ &
$\mp i28.66 $ & $\pm28.66$\\
$\sigma(0,\pm)$    & $ 2.12$ & $ -35.29$ & $+ i26.56 $ &
$\pm i28.66 $ & $\pm28.66$\\
\hline
\end{tabular}
\caption{Production spin density matrix elements (in fb) of $Z$ for the SM and the coefficients of various couplings in each matrix element
for unpolarized beams
 at $\sqrt{s}=500\text{~GeV}$.}
\label{table:3}
\end{table}}

{
\renewcommand{\arraystretch}{1.3}
\begin{table}[H]
\centering
\begin{tabular}{|c|c|c|c|c|c|}
 \multicolumn{5}{c}{} \\
 \hline
  & SM & $\text{Re}~b_{z}$  & $\text{Im}~b_{z}$ &
     $\text{Re}~\tilde{b}_{z}$&$ \text{Im}~\tilde{b}_{z}$   \\
\hline
$\sigma(\pm,\pm)$   & $ 8.98 $ &$-261.96 $ &$0$
& $0$ &  $\pm 369.38$ \\
$\sigma(0,0)$       & $63.57$ &$-261.96 $&$0$ &  $0$ &
 $0$\\
$\sigma(\pm,\mp)$  & $4.49$   &$-130.98$ &$0$ &$\mp i121.38 $  &  $0$\\
$\sigma(\pm,0)$    & $ 18.21$ & $ -303.29$ & $-i228.25 $ &
$\mp i246.3  $ & $\pm246.3 $\\
$\sigma(0,\pm)$    & $ 18.21$ & $ -303.29$ & $+i228.25 $ &
$\pm i246.3 $ & $\pm246.3 $\\
\hline
\end{tabular}
\caption{Production spin density matrix elements (in fb) of $Z$ for the
SM and the coefficients of various couplings in each matrix element for
($P_{L}=-0.8,\bar{P}_{L}=0.3$)   at $\sqrt{s}=500\text{~GeV}$.}
\label{table:4}
\end{table}}

We next evaluate, using equations listed in the previous section, the
leptonic asymmetries. For the SM, these are listed in Tables \ref{table:5} and
\ref{table:6}, together with the total cross sections, respectively for
$\sqrt{s}=250$ GeV and $\sqrt{s}=500$ GeV.
Note that except $A_{x}$, $A_{x^{2}-y^{2}}$ and $A_{zz}$, all other asymmetries 
vanish in the SM. This is because the asymmetries $A_x, \, A_{x^2-y^2}$ and $A_{zz}$ are CP even and  T even and they can occur at tree level in the SM. The remaining asymmetries vanish in the SM because they are either CP even and T odd and hence need an absorptive part in the amplitude to be nonzero, or CP odd, and hence proportional to CP violating parameters which are absent in the SM at tree level. With anomalous couplings, the CP even and T even asymmetries which are nonvanishing in the SM will get an additional contribution from $\text{Re}~b_Z$.

In Tables \ref{table:7} and \ref{table:8} we list, respectively for c.m.
energies 250 and 500 GeV, the additional
contributions to the cross section and various asymmetries for unit
values of couplings, the coupling listed in each row being the only one
which contributes to the observable in that row.

\begin{table}[H]
\centering
\begin{tabular}{ |p{2.5cm}|p{2.5cm}|p{2.5cm}|}
 \multicolumn{3}{c}{} \\
 \hline

\vspace{0.01 cm}Observable  & $P_{L}=0\newline\bar{P}_{L}=0$ &
  $P_{L}=-0.8\newline\bar{P}_{L}=0.3$ \\
\hline
$\sigma$ (in fb)   & $  243.67 $  &$342.24$\\
$A_{x}$& $ -0.014$ &

$ -0.085$\\
$A_{x^{2}-y^{2}}$ & $0.092$&$0.092$\\
$A_{zz}$   & $ -0.05$ & $   -0.05$ \\
\hline
\end{tabular}
\caption{The total production cross section (in fb) of $Z$ and the 
non-zero angular asymmetries in the SM 
for unpolarized and polarized beams
at $\sqrt{s}=250\text{~GeV}$.}
\label{table:5}
\end{table}

\begin{table}[H]
\centering
\begin{tabular}{ |p{2.5cm}|p{2.5cm}|p{2.5cm}|}
 \multicolumn{3}{c}{} \\
 \hline
\vspace{0.01 cm} Observable  & $P_{L}=0\newline\bar{P}_{L}=0$ &
     $P_{L}=-0.8\newline\bar{P_{L}}=0.3$ \\
\hline
$\sigma$ (in fb)   & $ 58.04 $ &$81.56$\\
$A_{x}$& $-0.012$ &
$-0.071$\\
$A_{x^{2}-y^{2}}$ & $0.035$ & $0.035$\\
$A_{zz}$   & $ -0.251$ & $ -0.251$ \\
\hline
\end{tabular}
\caption{The total production cross section (in fb) of $Z$ and the 
non-zero angular asymmetries in the
SM for unpolarized and polarized 
beams at $\sqrt{s}=500\text{~GeV}$.}
\label{table:6}
\end{table}

\begin{table}[h]
\centering
\begin{tabular}{| p{2cm}| p{2cm}|p{2cm}|p{2.2cm}|}
 \multicolumn{4}{c}{} \\
 \hline
\vspace{0.01 cm} Observable &\vspace{0.01 cm} Coupling & $P_{L}=0 \newline \bar{P}_{L}=0$ &
   $P_{L}=-0.8\newline \bar{P}_{L}=0.3$ \\
\hline
$\sigma$ (in fb)  & $\text{Re}~b_{z}$  & $-1400.17$   & $-1966.55$
\\
$A_{x}$&    $\text{Re}~b_{z}$  & 
$-0.0022$&$-0.014$\\
$A_{y}$ & $\text{Re}~\tilde{b}_{z}$&   
 $-0.026$&$-0.158$\\
$A_{z}$   & $ \text{Im}~\tilde{b}_{z}$ &$-0.242$& $-0.242$\\
$A_{xy}$   & $\text{Re}~\tilde{b}_{z}$  &$+0.344$&$+0.344$\\
 $A_{yz}$   & $\text{Im}~b_{z}$  &   $+0.041$ &$+0.253$\\
 $A_{xz}$   & $ \text{Im}~\tilde{b}_{z}$  &   $ -0.073$ &$-0.448$\\
 $A_{x^{2}-y^{2}}$   & $\text{Re}~b_{z}$  &   $  -0.082$ &$-0.082$\\
 $A_{zz}$   & $\text{Re}~b_{z}$  &   $ -0.289 $ &$-0.289 $\\
\hline
\end{tabular}
\caption{Anomalous contribution to cross section (in fb) 
and angular asymmetries for unpolarized and 
polarized beams at $\sqrt{s}=250\text{~GeV}$ for unit values of the
relevant couplings.}
\label{table:7}
\end{table}

\begin{table}[H]
\centering
\begin{tabular}{| p{2cm}| p{2cm}|p{2cm}|p{2.2cm}|}
 \multicolumn{4}{c}{} \\
 \hline
\vspace{0.01 cm}Observable & \vspace{0.01 cm} Coupling &  $P_{L}=0\newline\bar{P}_{L}=0$ &
     $P_{L}=-0.8\newline\bar{P}_{L}=0.3$ \\
\hline
$\sigma$ (in fb)   & $\text{Re}~b_{z}$ & $-559.5$& $-785.88$\\
$A_{x}$&    $\text{Re}~b_{z}$  & $+0.081$&   $+0.497$\\
$A_{y}$ & $\text{Re}~\tilde{b}_{z}$&   $ -0.157 $ &
 $-0.958 $\\
$A_{z}$   & $ \text{Im}~\tilde{b}_{z}$ &   $  -0.668 $ &$-0.668$\\
 $A_{xy}$   & $\text{Re}~\tilde{b}_{z}$  &   $ +0.948 $ &$+0.948$\\
 
 $A_{yz}$   & $\text{Im}~b_{z}$  & $+0.412$  &$+2.521 $\\
 $A_{xz}$   & $ \text{Im}~\tilde{b}_{z}$  &$-0.444$&
 $-2.720$\\
 $A_{x^{2}-y^{2}}$   & $\text{Re}~b_{z}$  &$ -0.685$ &$-0.685$\\
 $A_{zz}$   & $\text{Re}~b_{z}$  &$ -2.421$ &$-2.421$\\
\hline
\end{tabular}
\caption{Anomalous contribution to cross section (in fb) 
and angular asymmetries for unpolarized and 
polarized beams at $\sqrt{s}=500\text{~GeV}$ for unit values of the
relevant couplings.}
\label{table:8}
\end{table}

To obtain the $1\sigma$ limit $C_{\rm{limit}}$ on an anomalous coupling $C$ using 
these observables  we use 

\begin{equation}\label{limit}
C_{\rm{limit}}=\frac{\sqrt{1-A^{2}_{SM}}}{\vert A-A_{SM}\vert}\frac{1}{\sqrt{\sigma_{SM}\mathcal{L}}}
\end{equation}
{where $\sigma_{SM}$ is the SM cross section for the process $e^+e^- \to Z^*H \to \mu^+\mu^-H$, $A$ is the asymmetry for unit value of the coupling $C$, and
$A_{SM}$ is the corresponding value in the SM. 
Similarly, the limit from the cross section may be obtained from}
\begin{equation}\label{limit-cs}
C_{\rm{limit}}=\frac{1}{\vert \sigma-\sigma_{\rm{SM}}\vert}\sqrt{\frac{\sigma_{\rm{SM}}}{\mathcal{L}}}.
\end{equation}
We evaluate the limit considering one coupling 
to be non-zero at a time and list it in Tables \ref{table:9} and \ref{table:10} for the two different values of the c.m. energy and  assuming no systematic uncertainty.
\begin{table}[H]
\centering
\begin{tabular}{|c|c|c|c|}
   \hline
& & \multicolumn{2}{|c|}{Limit ($\times 10^{-3}$) for } \\
 Observable &Coupling &  $P_{L}=0$ &
     $P_{L}=-0.8$ \\
&&$\bar{P}_{L}=0$ &
$\bar{P}_{L}=0.3$ \\
\hline
$\sigma$   & $\text{Re}~b_{z}$  & $1.36$&$1.15$\\
$A_{x}$& $\text{Re}~b_{z}$  & $3480$ & $478 $\\
$A_{y}$ & $\text{Re}~\tilde{b}_{z}$&   $ 303 $ &$41.7 $\\
$A_{z}$   & $ \text{Im}~\tilde{b}_{z}$&   $ 32.3  $&$27.2$\\
$A_{xy}$   & $\text{Re}~\tilde{b}_{z}$  &   $22.7  $&$19.2 $\\
 $A_{yz}$   & $\text{Im}~b_{z}$  &   $189 $ &$26.1 $\\
 $A_{xz}$   & $ \text{Im}~\tilde{b}_{z}$  &   $ 107 $&$14.7 $\\
 $A_{x^{2}-y^{2}}$   & $\text{Re}~b_{z}$  &   $  94.5$&$80.2 $\\
 $A_{zz}$   & $\text{Re}~b_{z}$  &   $ 26.8  $ & $22.8 $\\
\hline
\end{tabular}
\caption{$1 \sigma$ limit obtained from various leptonic asymmetries for 
unpolarized and  polarized  beams
at $\sqrt{s}=250\text{~GeV}$.}
\label{table:9}
\end{table}
\begin{table}[H]
\centering
\begin{tabular}{|c|c|c|c|}
   \hline
& & \multicolumn{2}{|c|}{Limit ($\times 10^{-3}$) for } \\
 Observable &Coupling &  $P_{L}=0$ &
     $P_{L}=-0.8$ \\
&&$\bar{P}_{L}=0$ &
$\bar{P}_{L}=0.3$ \\
\hline
$\sigma$   & $\text{Re}~b_{z}$  & $3.32$&$2.8$
\\
$A_{x}$&  $\text{Re}~b_{z}$  & $394$&$54.2$\\
$A_{y}$ & $\text{Re}~\tilde{b}_{z}$&   $ 204 $ &
 $28.2 $\\
$A_{z}$   & $ \text{Im}~\tilde{b}_{z}$ &   $ 47.9 $ &$40.4 $\\
$A_{xy}$   & $\text{Re}~\tilde{b}_{z}$ &   $33.7 $ &$28.5 $ \\
 $A_{yz}$   & $\text{Im}~b_{z}$  &   $77.7$ &$10.7$ \\
 $A_{xz}$   & $ \text{Im}~\tilde{b}_{z}$  &   $ 72.0$ &$9.93 $ \\
 $A_{x^{2}-y^{2}}$   & $\text{Re}~b_{z}$  &   $  46.7$  & $39.4 $ \\
 $A_{zz}$   & $\text{Re}~b_{z}$  &   $ 12.8 $ & $10.8 $ \\
\hline
\end{tabular}
\caption{$1 \sigma$ limit obtained from various leptonic asymmetries for 
unpolarized and  polarized  
beams at $\sqrt{s}=500\text{~GeV}$.}
\label{table:10}
\end{table}

It is observed that the limit on the coupling $\text{Re}~b_{z}$ can be obtained 
from the observables $A_{x}$, $A_{zz}$ and $A_{x^{2}-y^{2}}$. 
However, among these we find that the observable $A_{zz}$ and the total 
cross section provide the best limits on the coupling $\text{Re}~b_{z}$, which 
becomes more stringent for the combination $(P_{L}=-0.8,\bar{P_{L}}=0.3)$ at 
$\sqrt{s}=500\text{~GeV}$. Also, we note that limits on the same obtained from the 
remaining observables slightly improves as one increases the c.m. energy 
especially for the combination where the electron polarization takes the 
negative sign. Although it seems that the total cross section is  enough for 
probing the coupling $\text{Re}~b_{z}$, one will require  the  other angular 
asymmetries to explore the couplings which do not appear in the total cross 
section.

{The best limit on $ \text{Im}~\tilde{b}_{z}$ is $9.93\times 10^{-3}$, 
which comes from  $A_{xz}$ whereas the best limit of $19.2 \times 10^{-3}$
on $\text{Re}~\tilde{b}_{z}$ can be obtained from the 
observable $A_{xy}$ for a reduced beam energy of 250 GeV. Similarly the best bound of $10.7 \times 10^{-3}$ on the coupling $\text{Im}~b_{z}$ can be achieved from the observable $A_{yz}$, 
which gets improved as one increases the c.m. energy to 500 GeV. As can be seen from the tables, the use of opposite sign beam polarization puts stronger constraints on the anomalous couplings, in some cases upto an order of magnitude better.}

{To estimate the effect of
systematic uncertainties, we have also evaluated the limits taking a
systematic error of 1\% for the asymmetries and the cross section.} We find  
that limits on couplings from various asymmetries worsen by a factor lying 
between 1.5 and 3 in the case of  c.m. energy of 250 GeV. They are thus in the same
ball park as the limits estimated in the absence of systematic uncertainty. 
However, the change in the limits for the case of c.m energy of 500 GeV is much 
smaller, only around 5-10\%. Similarly, {a 1\%  uncertainty}  in the 
measurement of the  cross section leads to a  change in the limits for the cross sections by 5-7\% at 500 GeV, {for unpolarized and polarized beams respectively}, whereas the {corresponding} limits worsen by a factor between 1.6 to 1.8 for the case of 250 GeV.

\section{Conclusions and discussion}\label{sec5}
 After the discovery of the Higgs boson, it is of utmost importance to have a 
precise measurement of its couplings 
with all other SM particles. We propose to measure the form and magnitude of the 
couplings of the Higgs boson to a pair of $Z$ bosons at an electron-positron collider 
by making use of the polarization data of the $Z$ boson. We adopt the formalism 
which connects angular asymmetries of charged leptons from $Z$ decay to the  polarization parameters of the $Z$. We take into account possible combinations of longitudinal beam 
polarizations likely to be available at the collider and have obtained the 
sensitivities of these polarization observables.

 We see that most of the $1\sigma$ limits are of the order of a few times 
$10^{-3}$.
Longitudinal polarization of the beam helps in increasing the sensitivities of certain 
observables. We find that that oppositely polarized beams provides tighter bounds on the couplings than the same sign polarized and unpolarized beams. Particularly a beam with $80\%$ electron polarization and $30\%$ 
positron polarization with opposite sign is found to be  capable of placing 
better limits on the anomalous couplings. Limits get slightly improved as one 
increases the c.m. energy. 

We have obtained the $Z$ polarization parameters and the asymmetries using the spin density matrix calculated at production level. However, at a collider these asymmetries will be determined from the angular distribution of the leptons from $Z$ decay and will thus entail acceptance and isolation cuts on the leptons. To get some idea of the cuts, we evaluated the asymmetries and sensitivities considered with generic acceptance cuts at the ILC: a 10 GeV cut on the lepton energy and a $5^{\circ}$ polar angle cut to keep away from the beam pipe. We find that these cuts lead to a less than $1\%$ change in all the observables including the total cross section.

We have also not considered Higgs decays, which do not affect the polarization parameters and asymmetries of the $Z$. The effect of Higgs decay on the sensitivities can be estimated by multiplying the SM cross section by the Higgs branching ratio and detection efficiencies in Eqns (\ref{limit}) and (\ref{limit-cs}). A full scale analysis using an event generator coupled with all appropriate cuts and detection efficiencies relevant to the decay channels of the $Z$ and Higgs with be able to refine the actual sensitivities that we have obtained.

\vskip .2cm
\noindent {\bf Acknowledgement}: KR acknowledges support from IIT Bombay, grant no. 12IRCCSG032. SDR acknowledges support from the Department of
Science and Technology, India, under the J.C. Bose National
Fellowship programme, Grant No. SR/SB/JCB-42/2009.


\end{document}